# High-precision molecular dynamics simulation of $UO_2$–$PuO_2$: superionic transition in uranium dioxide

S.I. Potashnikov[a], A.S. Boyarchenkov[a], K.A. Nekrasov[a], A.Ya. Kupryazhkin[a]

[a] Ural Federal University, 620002, Mira str. 19, Yekaterinburg, Russia

potashnikov@gmail.com  boyarchenkov@gmail.com  kirillnkr@mail.ru  kupr@dpt.ustu.ru

**Abstract**

Our series of articles is devoted to high-precision molecular dynamics simulation of mixed actinide-oxide (MOX) fuel in the rigid ions approximation using high-performance graphics processors (GPU). In this article we assess the 10 most relevant interatomic sets of pair potential (SPP) by reproduction of the Bredig superionic phase transition (anion sublattice premelting) in uranium dioxide. The measurements carried out in a wide temperature range from 300K up to melting point with 1K accuracy allowed reliable detection of this phase transition with each SPP. The λ-peaks obtained are smoother and wider than it was assumed previously. In addition, for the first time a pressure dependence of the λ-peak characteristics was measured, in a range from –5 GPa to 5 GPa its amplitudes had parabolic plot and temperatures had linear (that is similar to the Clausius-Clapeyron equation for melting temperature).

## 1. Introduction

Ionic crystals of uranium dioxide ($UO_2$) have a face-centered cubic (FCC) fluorite structure. Due to 15-fold difference in the atomic masses the oxygen sublattice is less stable (more mobile) than uranium sublattice, so at sufficiently high temperatures it melts down, and the anions move freely through the ordered structure of the cations. Such a phase transition accompanied by rapid increase of ionic conductivity is called superionic and observed in many other diatomic crystals. For $UO_2$ it was discovered by Dworkin and Bredig in 1967 [1] near 2700K from the temperature dependence curvature of enthalpy. Later, presence of the superionic transition was confirmed in other experiments [2] [3] [4]. Superionic transition appears in the form of λ-peak on the temperature dependence of heat capacity. Unfortunately, for $UO_2$ it is insufficiently investigated in experiments – only up to 2700K, while at higher temperatures different reviews give conflicting hypothetical recommendations [5] [6].

Molecular dynamics (MD) simulation has been already used to study superionic transition in $UO_2$ by some authors [7] [8] [9] [10]. However, all previous measurements were performed only for one set of pair potentials (SPP) and with low accuracy (charts have a small number of points with a big step in temperature). Unfortunately, this phase transition was not discussed in the review of Govers et al. [11], and a recent article devoted to energetic recoil simulation [12] mentions it just briefly. Authors of the latter have attempted to determine temperature of superionic transition by the chart of anion self-diffusion coefficient without logarithmic scale, which led them to incorrect conclusions. In particular, they claim that SPP Morelon-03 [13] have the lowest temperature of oxygen sublattice premelting.

However, our high-precision MD-methodology allowed reliable detection of the λ-peak characteristics (temperature and amplitude) for each of the 10 SPPs considered and demonstrated different results for superionic transition reproduction in $UO_2$ by them.

Also, in the previous work [14] we have revealed the discrepancy of model and experimental curves of heat capacity after the superionic transition region. This deviation may be caused by several reasons: divergence of model and experimental temperature dependences of the lattice constant, enthalpy and their derivatives, difference in the processes of anion sublattice disordering or inaccuracy of the empirical equations used to calculate the contribution of electronic and cationic defects to heat capacity. In order to clarify the situation, in this work we perform the corresponding analysis of various components of model heat capacity.

In addition, for the first time we have studied the pressure dependence of λ-peak characteristics in a wide range from –5 to 5 GPa.

## 2. Methodology

In this study, as in the previous one [14], we examined 10 the most relevant empirical SPPs for $UO_2$ in the rigid ions approximation: Walker-81 [15], Busker-02 [16], Nekrasov-08 [17], Goel-08 [18], obtained in the harmonic approximation from elastic properties at zero temperature; Morelon-03 [13], obtained using the lattice statics from energy of Frenkel and Schottky defects; Yamada-00 [7], Basak-03 [19], Arima-05 [20], Yakub-09 [21] and MOX-07 [10], obtained using MD simulation of thermal expansion and bulk modulus.

However, the results for Yamada-00 SPP are not discussed in this study, since its oxygen (anion) sublattice is unstable in the temperature range of 2050–2750 K. Namely, its lattice constant (L), enthalpy (H) and bulk modulus (B) undergo strong oscillations, so their numerical differentiation in this interval is meaningless. On the contrary, with the rest of SPPs they vary continuously with temperature and their derivatives, such as linear thermal expansion coefficient (LTEC) and isobaric heat capacity ($C_p$) have a smooth λ-peak.

Numerical parameters of the SPPs studied, as well as details of the measurement methodology of thermophysical properties (i.e. L, H, LTEC, B and $C_p$) may be found in the previous article [14].

In this work, all the MD simulations also were carried out on high-performance graphics processors (GPU) using NVIDIA CUDA technology, which gave us speedup of 100–1000 times (see details in [22] [23]). We used Ewald summation of electrostatic interactions in periodically-translated (i.e. without surfaces) cubic crystal of 4x4x4 FCC cells (768 particles). In order to integrate Newton's equations of motion we used Verlet method (with a time step of 5 fs) and Berendsen thermostat-barostat (with a relaxation time of 1 ps). The first 5 ps (1000 MD steps) of NPT-simulation (with a constant number of particles, temperature and pressure) were spent to reach equilibrium, and then observed quantities were averaged over the next 20 ps (4000 MD steps).

The isobaric heat capacity (Cp) was calculated by numerical differentiation of enthalpy which was measured at zero pressure (neglecting standard atmospheric pressure of ~0.1 MPa). However, in order to smooth curves on the charts with temperature step of 50K we averaged the enthalpy data over the interval of ±50K twice.

We measured superionic transition temperature ($T_\lambda$) with an accuracy of 50K from the λ-peak of the isochoric heat capacity ($C_v$) in order to exclude the thermal expansion influence. The pressure dependences of $T_\lambda$ for MOX-07 and Yakub-09 SPPs were measured with an accuracy of 5K and 100 ps simulations (instead of 25 ps).

Experimental data on the charts and in the tables are marked with a prefix "exp" and recommendations from reviews are marked with a prefix "rec".

## 3. Results and discussion

### 3.1. Analysis of the specific heat capacity components

Fig. 1 comprises isobaric heat capacity curves of the SPPs which, according to the previous study [14], have a pronounced λ-peak of superionic transition. These curves were drawn with the empirical contributions for electronic and cationic defects (small polarons and trivacancies) taken from the paper [24]: $C_e = 256*10790*\exp(-10790/T)/T^2$ kJ/(mol*K) and $C_c = 2*0.00000146*T$ kJ/(mol*K). Apart from them, there are two experimental dependences (data of Ralph [2] with λ-peak near 2610K and data of Ronchi [25] without the peak) and two corresponding recommendations (of Fink [5] with constant heat capacity above superionic transition and of IAEA with nonlinear growth extrapolation [6]). It is seen that Yakub-09 and MOX-07 SPPs coincide with the experimental dependences and the recommendations until 2600K while the rest of SPPs coincide only up to 1800K.

The discrepancy between model and experimental curves after the superionic transition observed in Fig. 1 may have several reasons: divergence of the corresponding temperature dependences of enthalpy and lattice constant (or their derivatives), difference in the processes of anion sublattice disordering or inaccuracy of the empirical approximations $C_e$ and $C_c$. In order to clarify this situation, in Fig. 2 we present the corresponding analysis of different contributions to the model heat capacity of MOX-07 SPP.

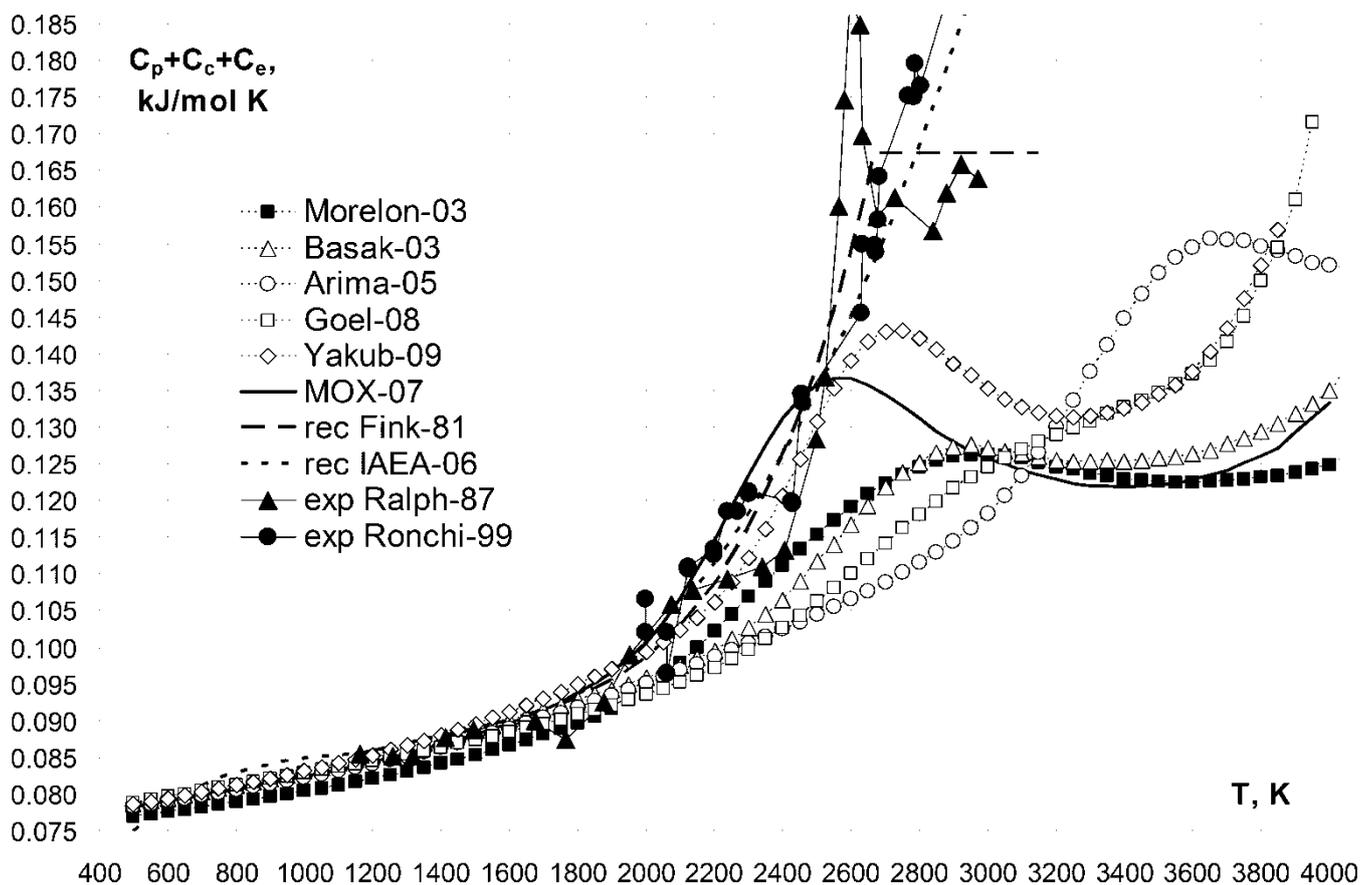

FIG. 1. Isobaric heat capacity with the empirical contributions $C_c(T)$ and $C_e(T)$.

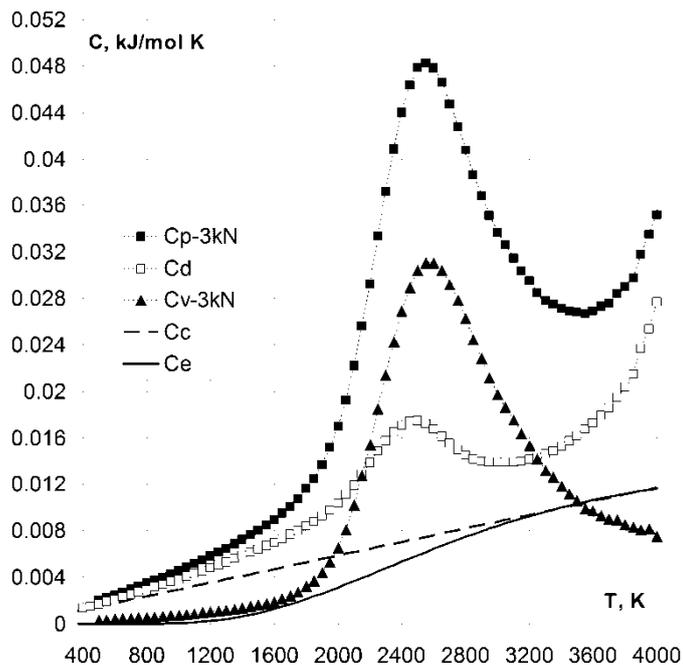

FIG. 2. Analysis of contributions to heat capacity by example of MOX-07 SPP.

At T < 2700K phonon vibrations of crystal lattice give the maximum contribution to heat capacity, which in the limit of high temperatures, according to the well-known models of Einstein and Debye [26], tends to the constant value of 3kN, where N is the number of particles in the crystal and k is the Boltzmann constant. Being normalized to one mole of $UO_2$ molecules, this constant equals to $9kN_A = 0.07483$ kJ/(mol*K), where $N_A$ is the Avogadro constant. Subtraction of this constant from calculated heat capacity values brings all contributions to the same scale allowing comparison of them.

At T < 1800K the principal contribution is a dilation term $C_d$, which is the difference between isobaric and isochoric heat capacities given by the well-known thermodynamic relation $C_d = C_p - C_v = 9\alpha^2 BTV_M$ (where α is the LTEC, B is the bulk modulus, T is the temperature, $V_M = N_A L^3/4$ is the volume of one mole of uranium dioxide molecules). Subtracting $C_d$ from $C_p$–3kN we obtain the term $C_v$–3kN, which is the contribution associated with defects formation. In the considered model it corresponds to anti-Frenkel disordering of the anion sublattice only.

From Fig. 2 it is evident that $C_v$–3kN sharply increases at T > 1800K up to the peak near 2600K and is greater than $C_d$ at 2150K < T < 3250K. However, in this range empirical terms $C_c$ and $C_e$ also become noticeable. Namely, the sum of them begins to exceed 0.01 kJ/(mol*K) and reaches 0.02 kJ/(mol*K) at 3150K.

Fig. 3 shows the comparison of $C_v$–3kN curves of various SPPs with the reference one named IAEA-06, which is derived from the IAEA recommendations [6] of L, LTEC, $C_p$ and the Martin's recommendation [27] of B. Note that on this chart λ-peaks are clearer than on any isobaric heat capacity charts, especially for Walker-81, Busker-02 and Nekrasov-08 SPPs (see their $C_p$ data in [14]).

The anionic sublattice disordering growth with temperature fully determines behavior of all the $C_v$–3kN curves obtained from MD simulations. Since every defect increases enthalpy, with saturation of defect concentration after superionic transition the curves should go down, exactly as observed in Fig. 3. However, the reference IAEA-06 curve (which includes contributions of electronic and cationic defects that are absent in the model curves) shows superlinear growth over all the range of temperatures. This fundamental difference may be due to underestimating of $C_e$ and $C_c$, since they were proposed more than 25 years ago [28], and sharper growth of these contributions could compensate the decrease of model curves after superionic transition.

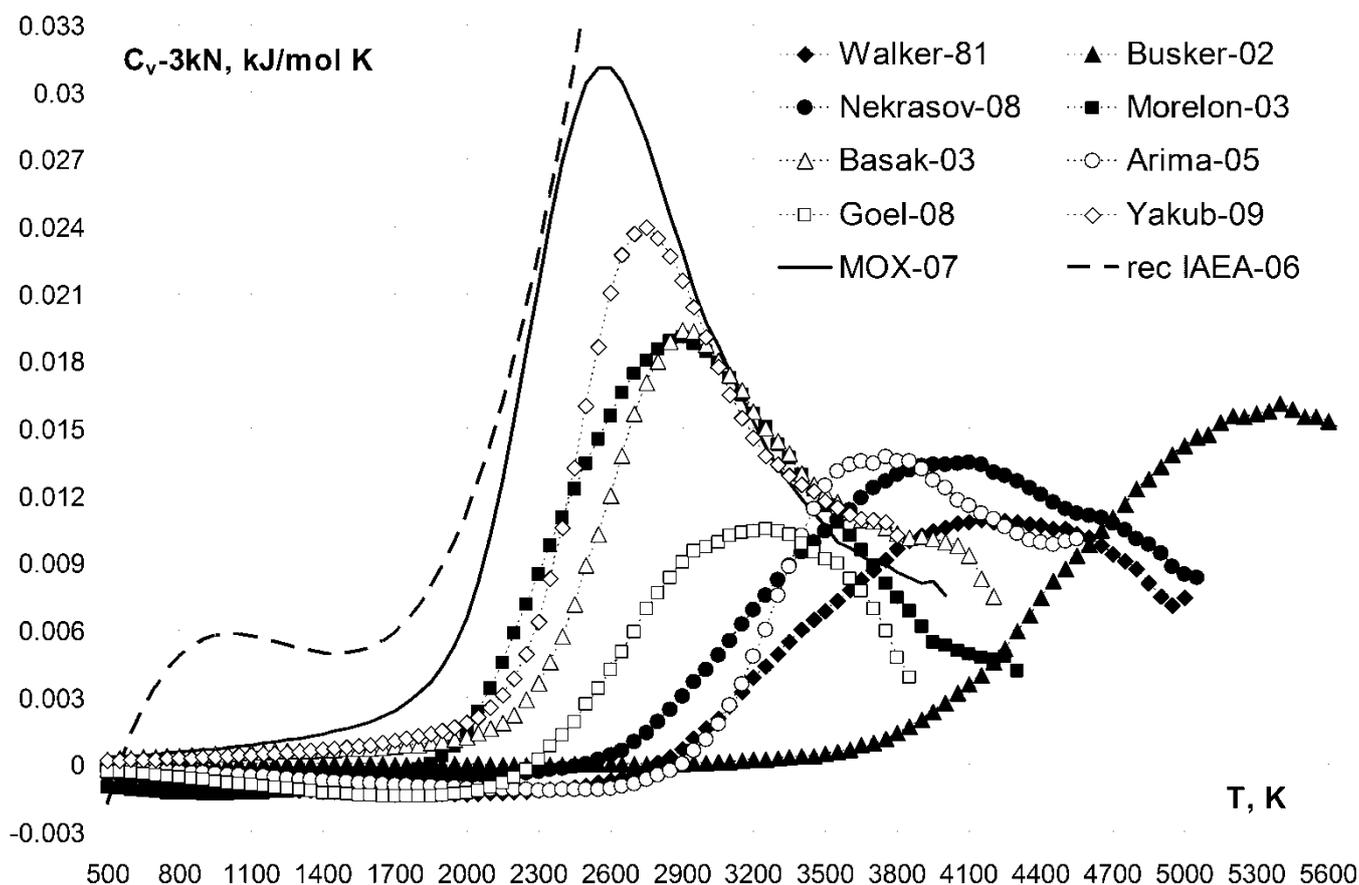

FIG. 3. Contribution of defect formation to heat capacity.

On the other hand, the divergence could be caused by inaccuracy of the experimental data for $C_p$ and $C_d$, difference of which determines the behavior of the reference $C_v$–$3kN$ curve. For example, for T > 1800K the IAEA review [6] gives uncertainty of the recommended data for $C_p$ as ±13% (i.e. 0.01–0.03 kJ/(mol*K)), but in the vicinity of superionic transition inapplicability of the equation and the uncertainty given is noted. In the most recent experimental measurements of high-temperature heat capacity of Ronchi et al. [25] the well-studied temperature range extends only up to 2700K, and accuracy of their technique did not allow detection of the λ-peak. Apart from them, the only known data in this range are Ralph's [2] with an error of ~0.02 kJ/(mol*K). According to these data, the heat capacity sharply decreases just after superionic transition near 2610K and then becomes constant. Thus, uncertainty of the recommendations above 2700K exceeds the difference between them and the model curves of MOX-07 and Yakub-09 SPPs.

Note also that all the model curves in Fig. 3 decrease after superionic transition until the corresponding melting point. Therefore, the growth of isobaric heat capacity after superionic transition seen in Fig. 1 is determined exclusively by the dilation term $C_d$. In particular, Arima-05 SPP has the maximum λ-peak value in Fig. 1 (because of the high LTEC peak [14]), but in Fig. 3 its λ-peak is 2.5 times lower than that of MOX-07.

### 3.2. Superionic transition: temperature, amplitude and pressure dependence of the λ-peak.

The chart of heat capacity in the review of Govers et al. [11] is drawn with a temperature step of 100K and is limited to a range of 0–3000K, so superionic transition could be detected only for a few SPPs, and λ-peak characteristics could not be determined reliably due to large scatter of values. On the contrary, in the present work measurements with precise step of 1K within a wide temperature range (from 300K up to melting point) allowed unambiguous characterization of superionic transition for all considered SPPs (except unstable Yamada-00 potentials).

Moreover, our MD simulation results demonstrated wide and smooth λ-peaks (see Fig. 1), with a range of temperatures over 1000K and amplitudes up to 0.03 kJ/(mol*K), indicating the diffuse phase transition with a continuous spectrum of equilibrium states (i.e. smooth increase of the anion sublattice disordering). This result fundamentally differs from the existing phenomenological hypotheses [3] [6] [9] of "delta-function" phase transition between the two equilibrium (crystalline and disordered) states of anionic sublattice with the corresponding narrow λ-peak (width of 30–50 K) of "infinite" amplitude.

The best reproduction of the quantitative superionic transition characteristics (in particular, the current IAEA recommendation of $T_\lambda$ = 2670K [6] and the Ralph's experimental value of $T_\lambda$ = 2610K [2]) is demonstrated by our MOX-07 potentials with $T_\lambda$ = 2600K and the maximum amplitude of λ-peak (see Table 1 and Fig. 3). Second to it are Yakub-09 with $T_\lambda$ = 2750K and 30% less amplitude of the peak, and the third are Basak-03 and Morelon-03 with $T_\lambda$ = 2900K and 60% less amplitude of the peak. Other SPPs considered in this work have $T_\lambda$ > 3250K (which exceeds the experimental melting point of $UO_2$) and their λ-peak amplitudes are 2–3 times lower than that of MOX-07.

The empirical rule of Bredig [29] for fluorite-type crystals states that $T_\lambda \approx 0.85 T_{melt}$. For $UO_2$ using the IAEA recommended melting temperature $T_{melt}$ = 3140±20K [6] this rule gives $T_\lambda \approx 2670$ K, which is in good agreement with the experimental values of Ralph [2] and Hiernaut et al. [3]. Our simulation results don't fit this rule, because of the high model meting temperatures [14]. To reach the coincidence, $T_{melt}$ for Morelon-03, Basak-03, Yakub-09 and MOX-07 potentials should be lower by 500–900K.

The model melting temperatures (independently of SPP) are significantly overestimated in comparison with the experimental values, due to the lack of surface (or other defects in the cationic sublattice) in MD simulations under periodic boundary conditions (PBC). So, the model crystals melt in a much superheated state (spinodal condition). In order to overcome this effect, some authors [11] [30] have measured the temperature of equilibrium of two-phase crystal-melt systems under PBC (binodal condition). But we believe that melting simulation of nanoscopic crystals with surface (which are finite and surrounded by vacuum, i.e. under isolated boundary conditions) would be more correct, and have devoted a separate article to this issue.

The achieved accuracy of measuring the λ-peak characteristics allowed us to obtain their pressure dependence. For the two SPPs Yakub-09 and MOX-07, which most adequately fit the experimental properties of uranium dioxide, we have measured the temperature dependences of thermophysical quantities at pressures ranging from –5 to 5 GPa. As an example of these curves, we provide the heat capacity dependences of MOX-07 in

Fig. 4 and the resulting pressure dependences for λ-peak amplitude and temperature in Fig. 5–6. One can note that $T_\lambda$ is much less dependent on pressure than $T_{melt}$ (displacement of 200–300K instead of 1500–2000K).

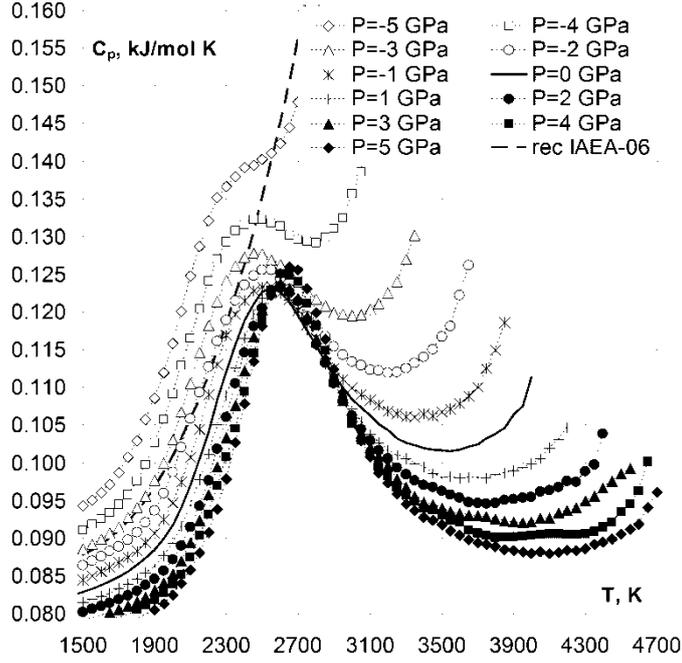

FIG. 4. Isobaric heat capacity of MOX-07 SPP under different pressures.

Due to the convergence of phase transitions temperatures $T_{melt}$ and $T_\lambda$ at negative pressures, the λ-peak in Fig. 4 becomes plateau at a pressure of –4 GPa, and almost disappears at –5 GPa, i.e. temperature dependence of heat capacity becomes non-decreasing. A similar behavior, but even in case of zero pressure, is demonstrated by Goel-08 potentials, which have the minimum difference of ~600K between $T_{melt}$ and $T_\lambda$ (see Fig. 1). Perhaps, the recent experiments on the specific heat measurements [25] have missed λ-peak just because of its small amplitude coupled with the closeness of the experimental temperatures $T_\lambda$ = 2670K and $T_{melt}$ = 3140K.

Pressure dependence of $T_{melt}$ and $T_\lambda$ explains the much lower estimates obtained for them in the study of Kupryazhkin et al. [8] on oxygen self-diffusion in $UO_2$ with Walker-81 potentials: $T_\lambda$ = 2700K instead of 4200K and $T_{melt}$ = 3200K instead of 5000K. In that work MD simulations under PBC were performed without barostat (i.e. NVT ensemble) and with the IAEA recommended lattice constant, which is by 0.2Å higher than the equilibrium lattice constant for these potentials (see [14]). Such straining of model crystal corresponds to a huge negative pressure (about –10 GPa) and leads to a significant decrease in $T_\lambda$ and $T_{melt}$ due to a rapid acceleration in the processes of lattice disordering, which is demonstrated in Fig. 5 by parabolic pressure dependence of λ-peak amplitude (fully determined by $C_V$–3kN contribution of defects to heat capacity).

The differential equation of Clausius-Clapeyron for pressure dependence of phase transition temperature, which is often used in the linear finite-difference form, is usually applied only to first-order transitions (in particular, to melting–crystallization). However, in our simulations the pressure dependence of $T_\lambda$ is also linear (see Fig. 6). The resulting linear equations for $T_\lambda(P)$ dependence, which are constructed independently from both the peaks of isobaric and isochoric heat capacities, are very similar for either Yakub-09 and MOX-07 SPPs, but the slope for Yakub-09 turned out to be 1.5–2 times steeper than for MOX-07.

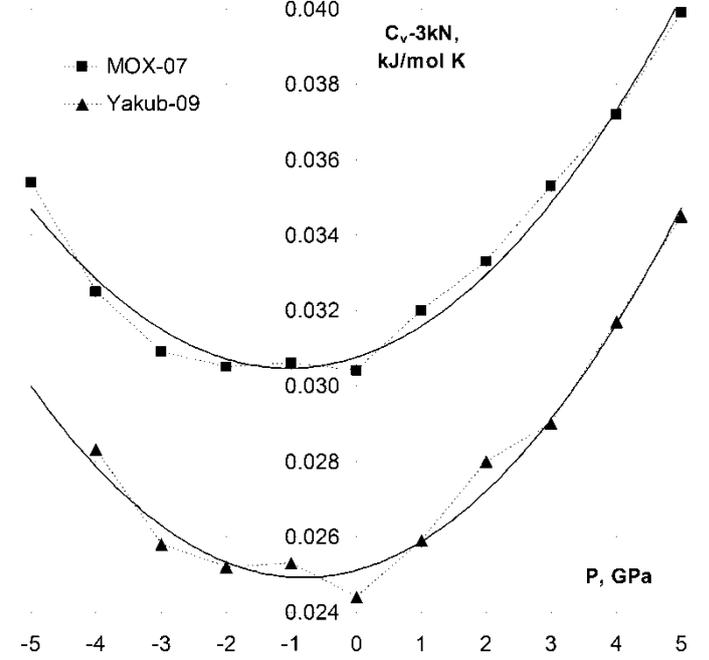

FIG. 5. Pressure dependence of λ-peak amplitude.

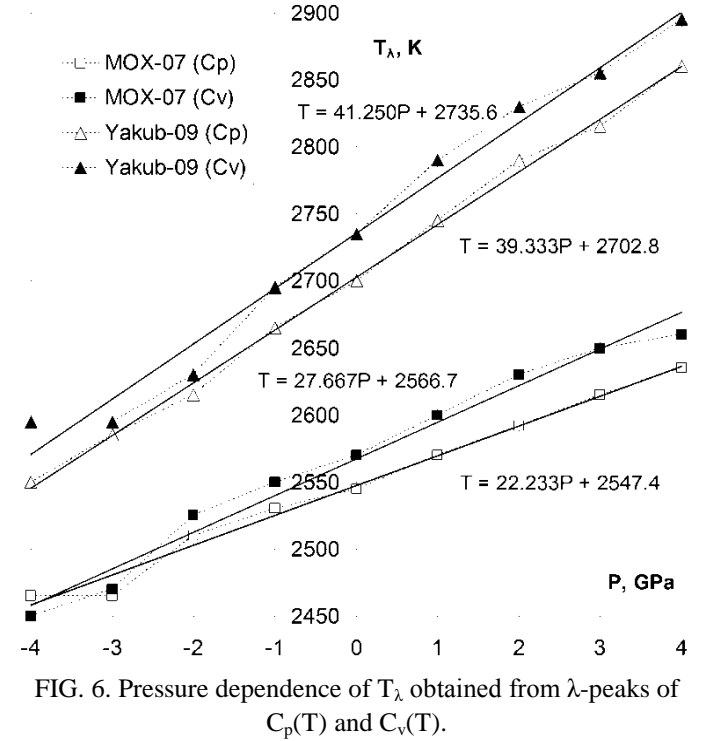

FIG. 6. Pressure dependence of $T_\lambda$ obtained from λ-peaks of $C_p(T)$ and $C_v(T)$.

## 4. Conclusion

Due to the high-precision measurements and analysis of contributions to heat capacity, we revealed the λ-peak of Bredig superionic phase transition with each of 10 considered sets of pair potentials (SPP) used for MD simulation of uranium dioxide ($UO_2$). Although λ-peak is not visible or unclear on a $C_p$ chart, it is unambiguously characterized on a $C_v$–3kN chart.

In contrast to the existing phenomenological hypotheses [3] [6] [9] of "delta-function" transition between the two (crystalline and disordered) equilibrium states of $UO_2$ anionic sublattice and the corresponding narrow λ-peak (of width 30–50 K) of "infinite" amplitude, we showed that MD simulation leads to wide and smooth λ-peaks with amplitude up to 0.03 kJ/(mol*K) and temperature range over 1000K. This indicates a diffuse phase transition with continuous spectrum of equilibrium states (corresponding to gradual disordering of anionic sublattice), which should be accompanied by gradual change of anti-Frenkel defect formation energy (see details in our next article on self-diffusion [31]).

The temperature dependence of $C_v$–3kN, which results from the isobaric heat capacity dependence by subtraction of the constant term of 3kN (phonon vibrations of the crystal lattice) and the dilation term $C_d = C_p$–$C_v$, is completely determined by anionic, cationic and electronic defects. In our MD simulations under periodic boundary conditions only anionic defects were modeled directly. Their contribution to heat capacity disappears just after complete disordering (premelting) of $UO_2$ oxygen sublattice, which gives clear λ-peaks and proper decrease of model $C_v$–3kN curves at the higher temperatures.

However, the behavior described is inconsistent with the current IAEA recommendation [6], which shows superlinear growth in the entire range of temperatures and lack of λ-peak at all. This indicates either that contribution of electronic and cationic defects should exceed the values calculated from the known empirical equations [28], or that there is significant inaccuracy in the experimental data at high temperatures.

The best reproduction of the quantitative superionic transition characteristics (in particular, the current IAEA recommendation of $T_λ$ = 2670K [6] and the Ralph's experimental value of $T_λ$ = 2610K [2]) is demonstrated by our MOX-07 potentials with $T_λ$ = 2600K and the maximum amplitude of λ-peak. Second to it are Yakub-09 with $T_λ$ = 2750K and 30% less amplitude of the peak, and the third are Basak-03 and Morelon-03 with $T_λ$ = 2900K and 60% less amplitude of the peak. Other SPPs considered in this work have $T_λ$ > 3250K (which exceeds the experimental melting point of $UO_2$) and their λ-peak amplitudes are 2–3 times lower than that of MOX-07.

Finally, for the first time pressure dependences of λ-peak characteristics were measured for MOX-07 and Yakub-09 potentials. It is shown that in a range from –5 to 5 GPa the λ-peak amplitude is a parabolic function of pressure, and $T_λ$ is a linear function similar to the Clausius-Clapeyron relation for melting temperature (though $T_λ$ varies only by 200–300K in the range of pressures where the corresponding melting temperature varies by 1500–2000K). The resulting linear equations for $T_λ(P)$ dependence, which are constructed independently from both the peaks of isobaric and isochoric heat capacities, are very similar for either SPPs, but the slope for Yakub-09 turned out to be 1.5–2 times steeper than for MOX-07.

## References

[1] A.S. Dworkin, M.A. Bredig, Journal of Physical Chemistry 72, 1277 (1968).
[2] J. Ralph, Journal of Chemical Society 83, 1253 (1987).
[3] J.P. Hiernaut, G.J. Hyland, C. Ronchi, International Journal of Thermophysics 14, 259 (1993).
[4] C. Ronchi, G.J. Hyland, Journal of Alloys and Compounds 213, 159 (1994).
[5] J.K. Fink, M.G. Chasanov, L. Leibowitz, Journal of Nuclear Materials 102, 17 (1981).
[6] Thermophysical Properties Database of Materials for Light Water Reactors and Heavy Water Reactors, IAEA (2006) http://www-pub.iaea.org/MTCD/publications/PDF/te_1496_web.pdf
[7] K. Yamada, K. Kurosaki, M. Uno et al., Journal of Alloys and Compounds 307, 1 (2000).
[8] A. Ya. Kupryazhkin, A.N. Zhiganov, D.V. Risovany, V.D. Risovany, V.N. Golovanov, Zh. Tekh. Fiz. 74, 114 (2004).
[9] E. Yakub, C. Ronchi, D. Staicu, Journal of Chemical Physics 127, 094508 (2007).
[10] S.I. Potashnikov, A.S. Boyarchenkov, K.A. Nekrasov, A.Ya. Kupryazhkin, ISJAEE 8, 43 (2007). http://isjaee.hydrogen.ru/pdf/AEE0807/AEE08-07_Potashnikov.pdf
[11] K. Govers, S. Lemehov, M. Hou, M. Verwerft, Journal of Nuclear Materials 376, 66 (2008).
[12] R. Devanathan, J. Yu, W.J. Weber, Journal of Chemical Physics 130, 174502 (2009).
[13] N.-D. Morelon, D. Ghaleb, J.-M. Dellhaye, L. Van Brutzel, Philosophical Magazine 83, 1533 (2003).
[14] S.I. Potashnikov, A.S. Boyarchenkov, K.A. Nekrasov, A.Ya. Kupryazhkin, High-precision molecular dynamics simulation of UO2–PuO2: pair potentials comparison, Journal of Nuclear Materials xxx (2011) xxx.
[15] J.R. Walker, C.R.A. Catlow, Journal of Physics C 14, 979 (1981).
[16] G. Busker, Ph.D. thesis, Imperial College, London (2002). http://busker.org/thesis/
[17] A.Ya. Kupryazhkin, A.N. Zhiganov, D.V. Risovany, K.A. Nekrasov et al., Journal of Nuclear Materials 372, 233 (2008).
[18] P. Goel, N. Choudhury, S.L. Chaplot, Journal of Nuclear Materials 377, 438 (2008).
[19] C.B. Basak, A.K. Sengupta, H.S. Kamath, Journal of Alloys and Compounds 360, 210 (2003).
[20] T. Arima, S. Yamasaki, Y. Inagaki, K. Idemitsu, Journal of Alloys and Compounds 400, 43 (2005).
[21] E. Yakub, C. Ronchi, D. Staicu, Journal of Nuclear Materials 389, 119 (2009).
[22] A.S. Boyarchenkov, S.I. Potashnikov, Numerical methods and programming 10, 9 (2009). http://num-meth.srcc.msu.ru/english/zhurnal/tom_2009/v10r102.html
[23] A.S. Boyarchenkov, S.I. Potashnikov, Numerical methods and programming 10, 158 (2009). http://num-meth.srcc.msu.ru/english/zhurnal/tom_2009/v10r119.html

[24] K. Kurosaki, K. Yamada, M. Uno et al., Journal of Nuclear Materials 294, 160 (2001).
[25] C. Ronchi, M. Sheindlin, M. Musella, G.J. Hyland, Journal of Applied Physics 85, 776 (1999).
[26] http://en.wikipedia.org/wiki/Debye_model
[27] D.G. Martin, High Temperatures High Pressures 21, 13 (1989).
[28] G.J. Hyland, J. Ralph, High Temperatures High Pressures 15, 179 (1983).
[29] M.A. Bredig, Proceedings of CNRS (France, 1971) 205, 183 (1972).
[30] T. Arima, K. Idemitsu, Y. Inagaki et al., Journal of Nuclear Materials 389, 149 (2009).
[31] S.I. Potashnikov, A.S. Boyarchenkov, K.A. Nekrasov, A.Ya. Kupryazhkin, High-precision molecular dynamics simulation of UO2–PuO2: anion self-diffusion in periodic crystals, Journal of Nuclear Materials xxx (2011) xxx.

TABLE 1. Bredig superionic transition temperature ($T_\lambda$) in Kelvins.

| Walker-81 | Busker-02 | Nekrasov-08 | Morelon-03 | Yamada-00 | Basak-03 | Arima-05 | Goel-08 | Yakub-09 | MOX-07 | exp Ralph-87 [2] | rec IAEA-06 [6] |
|---|---|---|---|---|---|---|---|---|---|---|---|
| 4200 | 5400 | 4050 | 2900 | 2100 | 2900 | 3750 | 3250 | 2750 | 2600 | 2610 | 2670 |